\documentclass[prb,twocolumn,showpacs]{revtex4}
\usepackage{graphicx}
\usepackage{amsmath}
\usepackage{amssymb}

\newcommand{\be}{\begin{equation}}
\newcommand{\ee}{\end{equation}}
\newcommand{\ba}{\begin{eqnarray}}
\newcommand{\ea}{\end{eqnarray}}

\begin{document}


\title{Unusual Microwave Response of Dirac Quasiparticles in Graphene}

\author{V.P.~Gusynin$^{1}$}
\author{S.G.~Sharapov$^{2}$}
\author{J.P.~Carbotte$^{2}$}

\affiliation{$^1$ Bogolyubov Institute for Theoretical Physics,
        Metrologicheskaya Str. 14-b, Kiev, 03143, Ukraine\\
        $^2$ Department of Physics and Astronomy, McMaster University,
        Hamilton, Ontario, Canada, L8S 4M1}

\date{\today }

\begin{abstract}
Recent experiments have proven that the  quasiparticles in graphene
obey a Dirac equation. Here we show that microwaves are an excellent
probe of their unusual dynamics. When the chemical potential is
small the intraband response can exhibit a cusp around zero
frequency $\Omega$ and this unusual lineshape changes to Drude-like
by increasing the chemical potential $|\mu|$, with width also
increasing linearly with $\mu$. The interband contribution at $T=0$
is a constant independent of $\Omega$ with a lower cutoff at $2
\mu$. Distinctly different behavior occurs if interaction-induced
phenomena in graphene cause an opening of a gap $\Delta$. At large
magnetic field $B$, the diagonal and Hall conductivities at small
$\Omega$ become independent of $B$ but remain nonzero and show
structure associated with the lowest Landau level. This occurs
because in the Dirac theory the energy of this level, $E_0 = \pm
\Delta$, is field independent in  sharp contrast to the conventional
case.
\end{abstract}




\maketitle

The band structure of graphene (a single atomic layer of graphite)
consists of two inequivalent pairs of the Dirac cones with apex at
the hexagonal Brillouin zone corners \cite{Wallace1947PRev}. For
zero chemical potential $\mu$ one cone of the pair is full and the
other empty. Through application of a gate voltage
\cite{Novoselov2004Science} the upper unoccupied cone can be
populated with electrons and by reversing the sign of the bias,
holes can be introduced in the lower energy cone. Since a $2+1$
dimensional Dirac equation governs the dynamics of quasiparticles in
graphene \cite{Semenoff1984PRL} many of its properties differ
significantly from those of other materials. The most striking
example is an unconventional quantization of the Hall conductivity
predicted in Refs.~\cite{Gusynin2005PRL,Peres2005}, unaware of the
experiments reported later in
Refs.~\cite{Geim2005Nature,Kim2005Nature}, which is related to the
anomalous properties of the lowest Landau level in Dirac theory.

Up to date all experimental studies (see e.g.
Refs.~\cite{Berger2004JPCB,Bunch2005Nano}) and most of the
theoretical work on graphene have been made for dc transport
properties. In this letter we study the microwave response of Dirac
quasiparticles and show that it has several anomalous properties
both in a strong magnetic field applied perpendicularly to the
graphene plane and in the absence of the field. We argue that an
experimental investigation of these features can provide added
support to the existing consensus that the carriers in graphene are
Dirac-like, but also shed light on not yet fully understood
questions such as the character of impurity scattering and the
presence of interaction induced phenomena. General expressions for
the real part of the longitudinal $\sigma_{xx}(\Omega)$ and Hall
$\sigma_{xy}(\Omega)$ conductivity for graphene at any temperature
$T$ and magnetic field $B$ are given in Ref.~\cite{Gusynin2005}.
They will not be repeated here although all numerical results
presented are based on these exact expressions. To interpret these
results we derive simple analytical expressions for
$\sigma_{xx}(\Omega)$ and $\sigma_{xy}(\Omega)$ that allow us to
elucidate the specifics of graphene.

\begin{figure}[h]
\centering{
\includegraphics[width=9cm]{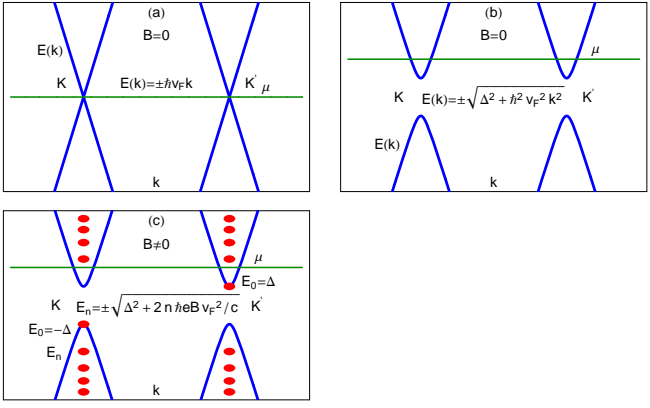}}
\caption{(Color online) Band structure of graphene. (a) The
low-energy linear-dispersion $E(\mathbf{k})$ near the Dirac
$\mathbf{K}$ and $\mathbf{K}^\prime$ points for $B=0$. (b) A
possible modification of the quasiparticle spectrum by the finite
gap (Dirac mass) $\Delta$ due to interaction-induced phenomena
\cite{Khveshchenko2001PRL,Gorbar2002PRB}. The  chemical potential
(indicated by horizontal line) $\mu$ is shifted from zero by the
gate voltage \cite{Novoselov2004Science}. (c) Landau levels $E_n$ in
the Dirac theory of graphene. For a given direction of the magnetic
field $\mathbf{B}$ applied perpendicularly to graphene's plane the
lowest $(n=0)$ Landau level has the energy $E_0=-\Delta$ at
$\mathbf{K}$ and $E_0=\Delta$ at $\mathbf{K}^\prime$. The presence
of a field independent $n=0$ level causes an unconventional Hall
effect and anomalies in the strong field microwave response.}
\label{fig:1}
\end{figure}

We begin with the simplest case of {\em massless Dirac
quasiparticles in zero field} which corresponds to the continuum
limit of noninteracting quasiparticles on a hexagonal lattice. The
corresponding band structure is shown in Fig.~\ref{fig:1}~(a).

In the limit of small impurity scattering rate $\Gamma(\omega)$
neglecting the real part of the impurity self-energy,
$\sigma_{xx}(\Omega,T)$ takes a particularly simple form
\begin{equation}
\label{B=0.cond-intra-inter}
\begin{split}
\sigma_{xx}(\Omega,T)&=\frac{e^2N_f}{2\pi^2
\hbar}\int\limits_{-\infty}^\infty
d\omega\frac{[n_F(\omega)-n_F(\omega^\prime)]}{\Omega}\frac{\pi}{4\omega\omega^\prime}\\
&\times\left[\frac{2\Gamma(\omega)}{\Omega^2+4\Gamma^2(\omega)}-\frac{2\Gamma(\omega)}
{(\omega+\omega^\prime)^2+4\Gamma^2(\omega)}\right]\\
&\times (|\omega|+|\omega^\prime|)(\omega^2+{\omega^\prime}^2),
\qquad \omega^\prime = \omega + \Omega,
\end{split}
\end{equation}
where $n_F(\omega)=1/[e^{(\omega-\mu)/T}+1]$ is the Fermi
distribution (we set $k_B=1$) and $N_f=2$ is the number of spin
components.  The first term of Eq.~(\ref{B=0.cond-intra-inter})
describes the intraband transitions and the second one the
interband. Similar expressions for the intraband contribution have
already appeared in the literature on $d$-wave superconductivity
\cite{Hirschfeld1994PRB,Carbotte2004PRB} and for the interband case
in the studies on $d$-density waves
\cite{Yang2002PRB,Valenzuela2005PRB,Benfatto2005PRB}. An essential
feature of Eq.~(\ref{B=0.cond-intra-inter}) is that we kept the
energy dependence of $\Gamma(\omega)$. In deriving this equation we
have also assumed the small $\Omega$ limit, so that
$\Gamma(\omega^\prime) \simeq\Gamma(\omega)$ and for $\Omega \ll T$
the difference $[n_F(\omega)-n_F(\omega^\prime)]/\Omega$ can be
replaced by the derivative $-\partial n_F(\omega)/\partial \omega$.

Let us consider for a moment, only the intraband term of
Eq.~(\ref{B=0.cond-intra-inter})
\begin{equation}
\label{B=0.cond-intra}
\sigma_{xx}(\Omega,T)=\sigma_{00}\int\limits_{-\infty}^\infty
d\omega \left(-\frac{\partial\, n_F(\omega)}{\partial\omega}\right)
\frac{2\pi|\omega|\Gamma(\omega)}{\Omega^2+4\Gamma^2(\omega)},
\end{equation}
with $\sigma_{00}=e^2N_f/(2\pi^2 \hbar)$.  If we take $\mu=0$ and
assume $\Gamma(\omega)= \gamma_{00}+ \alpha |\omega|$ with a small
value of $\gamma_{00}$ as expected in Born approximation in the weak
impurity scattering limit, we can get an approximate analytical
expression
\begin{equation}
\label{B=0.cond-intra-mu=0} \sigma_{xx}(\Omega,T) \simeq \frac{\pi
\sigma_{00}}{2 \alpha}\left[1- \frac{\pi}{8
\alpha}\frac{\Omega}{T}\right], \qquad \gamma_{00}< \Omega \ll T.
\end{equation}
Such an equation has been recently used in
Ref.~\cite{Carbotte2004PRB} to explain a cusp like behavior of the
microwave conductivity observed \cite{Turner2003PRL} in high purity
samples of ortho II YBCO$_{6.5}$  and some numerical results for
graphene are presented in Ref.~\cite{Ando2002JPSJ}. This shows that
if a cusp like behavior is observed in graphene, this would indicate
that Born approximation is relevant. If, however, the chemical
potential is increased so that $|\mu|>T$
Eq.~(\ref{B=0.cond-intra-mu=0}) is replaced by
\begin{equation}
\label{B=0.cond-intra-mu} \sigma_{xx}(\Omega,T) = \sigma_{00} 2 \pi
|\mu| \frac{(\gamma_{00} + \alpha |\mu|)}{\Omega^2 + 4 (\gamma_{00}
+ \alpha |\mu|)^2}
\end{equation}
which has a Drude form. Here the factor of $|\mu|$ gives the amount
of spectral weight which is distributed according to a Drude form
with scattering rate $\gamma_{00}+ \alpha |\mu|$. The width of the
Drude peak can be increased simply by changing the gate voltage,
$\mu \varpropto \mbox{sgn}\,V_g \sqrt{|V_g|}$ with impurity
scattering left the same. What has happened is that $\Gamma(\omega)$
near $\omega=\mu$ rather than  near $\omega=0$ has now become the
relevant quantity and it increases with $|\mu|$. It is important to
note that the frequency dependence of $\Gamma(\omega)$ in the more
general case could be mapped out in this way, because the shape of
$\Gamma(\omega)$ gets reflected in the shape of
$\sigma_{xx}(\Omega)$. For example, in the unitary limit of a
$T$-matrix approximation the dependence of $\Gamma(\omega)$ is
completely different from Born limit and is inversely proportional
to the electron density of states \cite{Peres2005}. This dependence
in turn modulates the $\Omega$ dependence of the intraband term.

Next we return to Eq.~(\ref{B=0.cond-intra-inter}) and consider the
strict $\Gamma =0$ limit and $T=0$ when it reduces to (see
Ref.~\cite{Yang2002PRB})
\begin{equation}
\label{B=0.cond-intra-inter-T=0} \sigma_{xx}(\Omega)= \frac{\pi
e^2N_f}{h}|\mu|\delta(\Omega)+\frac{\pi
e^2N_f}{4h}\theta\left(\frac{|\Omega|}{2} -|\mu|\right).
\end{equation}
For $\mu=0$ there is no intraband Drude contribution here
proportional to $\delta(\Omega)$ because $\Gamma=0$. Also  the
interband contribution extends from zero to the band edge (assumed
very large in this work). It is independent of energy $\Omega$ and
in the bare bubble approximation used to derive
Eqs.~(\ref{B=0.cond-intra-inter}), (\ref{B=0.cond-intra-inter-T=0})
is equal to $\pi e^2N_f/(4h)$ in height \cite{Ando2002JPSJ}. The
flatness of this response can be traced to the topology of the Dirac
cone  and the linear in momentum dependence of the Dirac
quasiparticles. As $\mu$ increases, the interband transitions become
gapped at $\Omega = 2 |\mu|$ and the missing spectral weight
reappears in the $\delta(\Omega)$-function contribution.

A nonlinear behavior of the diagonal magnetoresistivity, $\rho_{xx}$
in Kish graphite in high magnetic field has been observed and
interpreted in Ref.~\cite{Iye1985PRL} in terms of the formation of
the new charge-density-wave phase. Also as predicted in
Refs.~\cite{Khveshchenko2001PRL,Gorbar2002PRB}, an
interaction-induced phenomena may cause an opening of an excitonic
gap (see Fig.~\ref{fig:1}~(b)) in the quasiparticle spectrum.
Depending on the model parameters, the gap opens in zero or in a
finite  magnetic field. Recent measurements \cite{Kopelevich2006PLA}
of the Hall resistivity made on highly oriented pyrolytic graphite
indicate that there is a gap at $B=0.03 \mbox{T}$.  In our opinion,
if the value of the gap is expected to be the order of a few Kelvin,
microwaves are the best technique to detect its existence. In this
instance, for {\em massive Dirac quasiparticles in zero field\/}
Eq.~(\ref{B=0.cond-intra-inter-T=0}) acquires the form
\begin{equation} \label{B=0.cond-Delta-T=0}
\begin{split}
\sigma_{xx}(\Omega)&=\frac{\pi
e^2N_f}{h}\delta(\Omega)\frac{(\mu^2-\Delta^2)\theta(\mu^2 -
\Delta^2)}{|\mu|} \\&+\frac{\pi
e^2N_f}{4h}\frac{\Omega^2+4\Delta^2}{\Omega^2}\theta\left(\frac{|\Omega|}{2}
-{\rm max}(|\mu|,\Delta)\right).
\end{split}
\end{equation}
Here the theta function $\theta\left(|\Omega|/2 -{\rm
max}(|\mu|,\Delta)\right)$ cuts off the low $\Omega$ part of the
interband contribution at $2|\mu|$ or $2\Delta$ whichever is the
largest. The additional factor $(\Omega^2 + 4 \Delta^2)/\Omega^2$ is
to be noted and allows one to distinguish in $\sigma(\Omega)$ the
case of finite $\Delta$ from finite $\mu$. As we saw in relation to
Eq.~(\ref{B=0.cond-intra-inter-T=0}) when $\Delta=0$ the interband
contribution remains flat although cuts off at $2|\mu|$. For finite
$\Delta$ this is no longer that case and for $|\mu| < \Delta$ the
interband spectral weight at the gap has been increased by a factor
of $2$. This is a signature of a gap opening and effectively
represents a change in the geometry of the Dirac cones as
illustrated in Fig.~\ref{fig:1} (from (a) to (b) panel).
\begin{figure}[h]
\centering{
\includegraphics[width=7cm]{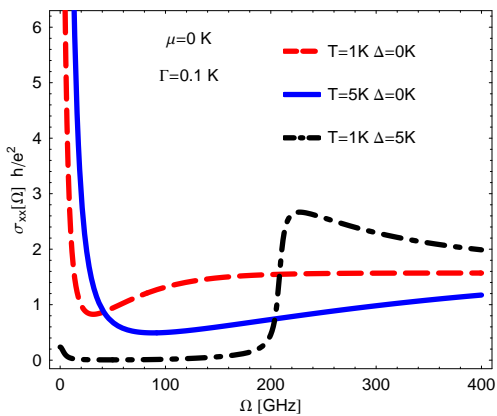}}
\caption{(Color online) The microwave conductivity
$\sigma_{xx}(\Omega,T)$ in units $e^2/h$ vs microwave frequency
$\Omega$ in GHz. The long dashed is for $T=1\mbox{K}$ and the solid
for $T=5\mbox{K}$. In both cases $\mu=\Delta=0$ and
$\Gamma=0.1\mbox{K}$. The dash-dotted curve has a gap
$\Delta=5\mbox{K}$ and $T=1\mbox{K}$.} \label{fig:2}
\end{figure}
In Fig.~\ref{fig:2} we show results for the microwave conductivity
$\sigma_{xx}(\Omega,T)$ vs $\Omega$ at finite temperature. The long
dashed curve for $T=1 \mbox{K}$ shows a Drude peak at small $\Omega$
(intra) superimposed on a interband contribution of $\pi e^2/(2h)$
constant independent of $\Omega$ except for some depletion below
$200 \mbox{GHz}$ due to finite temperature with optical spectral
weight transferred to the Drude peak. When $T$ is increased to $5
\mbox{K}$ (solid curve), the thermal depletion increased
considerably with increased weight in the Drude peak. On the other
hand, if instead of increasing $T$ we open a gap $\Delta=5 \mbox{K}$
(the dash dotted curve), the Drude peak is almost completely
depleted as is the region below $2\Delta \simeq 200\mbox{GHz}$ with
extra optical weight accumulating at and above $2\Delta$ in the
interband region, where it is distributed over the region of $\Omega
\gtrsim (2\div4) \Delta$. The peak at $\Omega = 2 \Delta$ in
$\sigma_{xx}(\Omega)$ is characteristic of the opening of a gap as
we have already described.

\begin{figure}[h]
\centering{
\includegraphics[width=8cm]{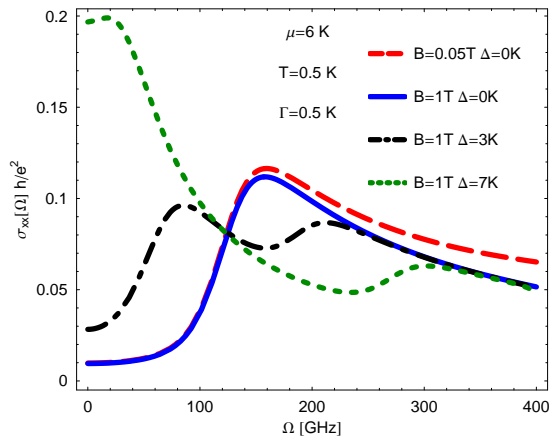}}
\caption{(Color online) Conductivity $\sigma_{xx}(\Omega,T)$ in
units $e^2/h$ vs microwave frequency $\Omega$ in GHz. The chemical
potential is set at $6\mbox{K}$, the temperature at $0.5 \mbox{K}$
and the impurity scattering rate assumed constant set at
$0.5\mbox{K}$. Long dashed curve $B=0.05 \mbox{T}$ and $\Delta=0$,
solid curve $\Delta=0$, dash-dotted $\Delta=3\mbox{K}$ and short
dashed $\Delta=7\mbox{K}$. All three last curves are for
$B=1\mbox{T}$.} \label{fig:3}
\end{figure}

{\em Massless and massive Dirac quasiparticles in a strong field.}
The latest measurements in graphene samples \cite{Zhang2006} show
that at $B=25 \mbox{T}$ near $\mu =0$ the diagonal and Hall dc
magneto-resistances do not conform to the standard quantum Hall
observation. Here we target the microwave response in the same
region. When a magnetic field $B$ is applied, the expressions for
the conductivity become very complicated as given in
Ref.~\cite{Gusynin2005}. In the very high field limit the Landau
levels with $n\neq0$ can all be shifted to high energy, but as we
have already  illustrated in Fig.~\ref{fig:1}~(c) this is not the
case for $n=0$ level. This level may only depend on the field
indirectly if the gap $\Delta(B,\mu)$ is induced by the field
\cite{Gorbar2002PRB} (the phenomenon of magnetic catalysis
\cite{Gusynin1995PRD}), so that it remains well below the rest of
the levels. If we take $B$ very large, only the contribution $\sim
1/B$ associated with the transitions from $n=0$ to $n=1$ levels
survives and after multiplying it by the degeneracy factor of the
Landau levels $\sim B$, we arrive at a very simple formula
\begin{equation} \label{B.cond-Delta}
\begin{split}
&\sigma_{xx} (\Omega,T)  = \frac{e^2N_f\Gamma}{2\pi h\,\Omega}{\rm
Im}\left[\Psi\left( \frac{\Gamma+i(\mu+\Omega+\Delta)}{2\pi
T}+\frac{1}{2}\right) \right.\\
&- \left.\Psi\left(\frac{\Gamma+i(\mu-\Omega+\Delta)}{2\pi
T}+\frac{1}{2}\right) + (\Delta \to - \Delta)\right],
\end{split}
\end{equation}
where $\psi$  is the digamma function. Here for simplicity we
neglected the Zeeman splitting in the strong field.
Eq.~(\ref{B.cond-Delta}) which is applicable for an arbitrary
relationship between $\Omega,T$ and $\Gamma$, shows that the
diagonal conductivity $\sigma_{xx}(\Omega,T)$ is independent of $B$,
but the position in energy of the two inequivalent $n=0$ Landau
orbitals namely $\mu-\Delta(B,\mu)$ and $\mu+\Delta(B,\mu)$ measured
relative to the chemical potential (see Fig.~\ref{fig:1}~(c))
remains in the problem. Results based on calculation of the full
expression for $\sigma_{xx}(\Omega,T)$ from Ref.~\cite{Gusynin2005}
are shown in Fig.~\ref{fig:3} and we have verified that they are in
a good agreement with the approximate expression
(\ref{B.cond-Delta}).
There are four curves. Comparison of solid and long-dashed curves
shows that the large field limit is already attained for
$B=0.05\mbox{T}$. Here the gap $\Delta=0$ and the chemical potential
equal to $6\mbox{K}$ is clearly identifiable at $120\mbox{GHz}$,
where a large rise smeared by temperature $($0.5 \mbox{K}$)$ and
impurities $\Gamma=0.5 \mbox{K}$ is seen in $\sigma_{xx}$. When a
gap develops, however, the shape of the curve is completely changed.
In both long-short dashed and short dashed curves, structure at $\mu
-\Delta$ and $\mu+\Delta$ is clearly visible reflecting that the
energy of the $n=0$ Landau level at $\mathbf{K}$ point is $\Delta$
and at $\mathbf{K}^\prime$ point is $-\Delta$, respectively. This is
yet another clear signature of the Dirac character of the
quasiparticles in graphene which can be used to detect the gap.
Indeed since in large fields all changes in the shape of
$\sigma_{xx}(\Omega)$ are caused by the gap, one may consider that
if the gap $\Delta(B)$ develops as the field increases, the curves
with $\Delta=0$, $\Delta=3\mbox{K}$ and $\Delta=7\mbox{K}$ shown in
Fig.~\ref{fig:3} would actually show up as the field increases.

Finally we consider the microwave Hall conductivity which can also
provide more insight into the properties of graphene. It can be
shown that under the same conditions as were used to derive
Eq.~(\ref{B.cond-Delta}), the microwave Hall conductivity is given
by
\begin{equation}
\label{B.Hallcond-Delta}
\begin{split}
\sigma_{xy} &(\Omega)=- \frac{e^2N_f{\rm sgn}(eB)}{\pi h}\left\{{\rm
Im} \left[\Psi\left(\frac{\Gamma+i(\mu+\Delta)}{2\pi
T}+\frac{1}{2}\right)\right. \right.\\
&  + (\Delta \to - \Delta)\Big] + \frac{\Gamma}{2\Omega}{\rm Re}
\left[\Psi\left(\frac{\Gamma+i(\mu+\Omega+\Delta)}{2\pi
T}+\frac{1}{2}\right)\right. \\
& \left. - \Psi\left(\frac{\Gamma+i(\mu-\Omega+\Delta)}{2\pi
T}+\frac{1}{2}\right) + (\Delta \to - \Delta) \right]\Bigg\}.
\end{split}
\end{equation}
\begin{figure}[h]
\centering{
\includegraphics[width=7cm]{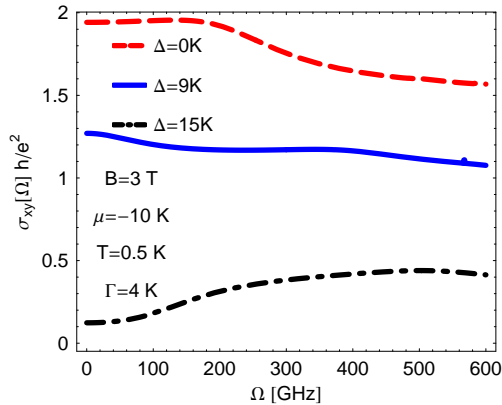}}
\caption{(Color online) Conductivity $\sigma_{xy}(\Omega,T)$ in
units $e^2/h$ vs microwave frequency $\Omega$ in GHz. The chemical
potential is set at $-10\mbox{K}$, the temperature at $0.5 \mbox{K}$
and the impurity scattering rate assumed constant set at
$4\mbox{K}$. The gap $\Delta$ is also indicated in the figure.}
\label{fig:4}
\end{figure}
In Fig.~\ref{fig:4} we show results for the frequency dependence of
the Hall conductivity in units of $e^2/h$. As found in
Ref.~\cite{Gusynin2005} the opening of a gap decreases the dc value
of the Hall conductivity. For the case $\Delta=0$ (dashed line)
there is additionally structure at $\Omega \approx |\mu|$, while for
$\Delta$ finite the equivalent structure is moved to energies
$|\mu|+\Delta$ and $|\mu|-\Delta$ (see Fig.~\ref{fig:1}~(c)) as we
have seen in the case of diagonal conductivity.

Up to now we have considered the influence of the gap opening on the
conductivity  when the gap itself was set to a constant. A more
interesting possibility is to explore the dependence of the gap
$\Delta$ on $\mu$ and $B$ by changing the gate voltage or the
magnetic field. It is expected that a decrease in $\mu$ or an
increase of $B$ favors the opening of the  gap when $|\mu|<\mu_c$ or
$B>B_c$ as was predicted in Ref.~\cite{Gorbar2002PRB}. The important
conclusion of our studies is that the generation of an excitonic gap
should lead to a new insulator phase around the point $\mu=0$ or in
very large magnetic fields  and this would be one more striking
difference of QHE in graphene from standard semiconductors where the
fractional quantum Hall effect is instead observed when entering the
lowest Landau level.

In summary the microwave response of graphene shows many distinct
characteristics which reflect the Dirac nature of the
quasiparticles. These include intra and interband contributions to
the conductivity in zero magnetic field and microwave response at
large fields. While in conventional case large $B$ would shift all
Landau levels to high energies above the microwave region for the
Dirac case the lowest Landau level contribution remains unshifted
and affects $\sigma_{xx}(\Omega)$  and $\sigma_{xy}(\Omega)$ in the
small $\Omega$ region in a distinct way which can be used to study
the effect of a gap opening induced by the magnetic field.

We thank Yu.G.~Pogorelov for useful discussion. The work of V.P.G.
was supported by the SCOPES-project IB7320-110848 of the Swiss NSF.
J.P.C. and S.G.Sh. were supported by the Natural Science and
Engineering Research Council of Canada (NSERC) and by the Canadian
Institute for Advanced Research (CIAR).

\end{document}